\documentclass{INTERSPEECH2023}

\interspeechcameraready

\title{Language-Universal Phonetic Representation in Multilingual Speech Pretraining for Low-Resource Speech Recognition}
\name{Siyuan Feng,
Ming Tu, Rui Xia, Chuanzeng Huang, Yuxuan Wang}
\address{
  Speech and Music Intelligence (SAMI), ByteDance}
\email{\{fengsiyuan.ee,mingtu,rui.xia,huangchuanzeng,wangyuxuan.11\}@bytedance.com}

\begin{document}

\maketitle
 
\begin{abstract}
We improve low-resource ASR by integrating the ideas of multilingual training and self-supervised learning. Concretely, we leverage an International Phonetic Alphabet (IPA) multilingual model to create frame-level pseudo labels for unlabeled speech, and use these pseudo labels to guide hidden-unit BERT (HuBERT) based  speech pretraining in a phonetically-informed manner. The experiments on the Multilingual Speech (MLS) Corpus show  that the proposed approach consistently outperforms the standard HuBERT  on all the target languages. Moreover, on 3 of the 4 languages, comparing to the standard HuBERT, the approach performs better, meanwhile is able to save  supervised training data by 1.5k hours  (75\%)  at most. Our approach outperforms most of the state of the arts, with much less pretraining data in terms of hours and language diversity.  Compared to XLSR-53 and a retraining based multilingual method, our approach performs better with full and limited finetuning data scenarios.

\end{abstract}

\section{Introduction}
\label{sec:intro}
Recent advancement of end-to-end (E2E) automatic speech recognition (ASR) has been witnessed in high-resource languages e.g. English and Mandarin. One driving force behind is the availability of large amounts of speech data resources.  E2E ASR models are data-hungry \cite{li2022recent}. Most of the approximately 7000 languages in the world  are considered under-resourced \cite{speech2020scharenborg}, in the senses of transcribed and/or untranscribed speech data. 
Transcribing speech is time-consuming,  expensive, and sometimes  impossible e.g. for languages with no orthography.
This poses a major challenge in developing high-performance ASR systems for low-resource languages.


To tackle the low-resource ASR task, 
past studies proposed various categories of approaches. 
one mainstream research line is multilingual training \cite{Huang2013cross,toshniwal2018multilingual}, which aims to  increase the training data amount and facilitate knowledge sharing among languages.  
A second research line is crosslingual transfer \cite{swietojanski2012unsupervised,hou2021exploiting}, transferring one or multiple non-target language's knowledge to the target language. 
Recently, the application of self-supervised learning (SSL)  in the low-resource ASR task has achieved remarkable performances \cite{oord2018cpc,baevski2020wav2vec,hsu2021hubert,chung2021w2v,baevski2022data2vec}. SSL  utilizes supervision information derived  from the input data, making it intrinsically suitable for limited-supervision scenarios. In SSL, contrastive learning \cite{baevski2020wav2vec} and masked prediction \cite{devlin2018bert}
are two commonly adopted strategies:  contrastive learning tries to distinguish a target sample  from distractor samples  given an anchor representation (e.g. wav2vec 2.0 \cite{baevski2020wav2vec}), while  masked prediction tries to output a distribution over
a discrete vocabulary given a masked input utterance (e.g. hidden-unit BERT (HuBERT) \cite{hsu2021hubert}). In addition to the aforementioned research lines, data augmentation and self-training,  which share the 
idea of increasing the supervised data amount, were studied 
\cite{meng2021mixspeech,zhang2021xlst}.


This study addresses the low-resource ASR task by integrating the ideas of  multilingual training and SSL. Specifically, we investigate effective and data-efficient  pretraining strategies to improve HuBERT \cite{hsu2021hubert}, encouraging the model to  explicitly learn language-universal phonetic representations through multilingual phonetic sharing. Multilingual phonetic sharing is motivated by the  analysis   \cite{Zelasko2020That_interspeech} which proved that phonetics of different languages are highly shareable, and such sharing  benefits multilingual phone recognition. Phonetic sharing can be realized by the language-independent International Phonetic Alphabet (IPA) system \cite{Zelasko2020That_interspeech,feng2021how_phonotactics}. 
In our proposed approach, we first train a multilingual,  hopefully language-universal IPA  model which recognizes  IPA symbol sequences from speech utterances. This single IPA model captures a broad range of languages' phonetic information. Next, we leverage the IPA model to create pseudo frame labels for multilingual untranscribed speech data. The IPA pseudo labels are used to conduct HuBERT pretraining instead of K-means based labels \cite{hsu2021hubert}. 
Compared to K-means based pseudo labels, IPA labels encode much richer phonetic knowledge. Guided by the phonetically-informed IPA labels, the pretrained model is capable of capturing a language-universal phonetic representation, a property that benefits downstream ASR finetuning, especially for a language with very limited finetuning data.

Building an IPA   model requires supervised data.
Our experiments will demonstrate that training of an IPA model is not data-hungry: 100 hours per language suffice. We will show in an ablation study  that good IPA pseudo labels can be inferred by an IPA model which is not fully optimized in terms of the phone recognition performance.
Furthermore, compared to a standard HuBERT, 
our proposed approach performs better in ASR tasks, 
and is able to save up to 1.5k hours  (75\%) of supervised finetuning data of some target language. In comparison to another multilingual method which is based on  IPA model retraining,
our approach performs better and is more robust to  limited finetuning data.
Notably, a recent study \cite{wang20222supervision} shared a similar idea by replacing K-means labels with phoneme alignments derived by a hybrid ASR model in HuBERT pretraining. Unlike IPA, the  phoneme system is language-dependent \cite{enwiki:1099218041}. Our approach is thus not limited to monolingual ASR as was discussed in \cite{wang20222supervision}.

\section{Proposed approach} 
\subsection{Language-universal IPA model as pseudo labeler}

Training a language-universal IPA model requires  $N$  languages' supervised training data (denoted as $\mathcal{F}$) and phonetic IPA transcripts. 
The acquisition of IPA transcripts is by conversion from orthographic transcripts leveraging a  Grapheme-to-IPA tool such as   \cite{hasegawa2020grapheme}.
The IPA symbols are language-independent, thus the $N$ languages' training data can be directly merged for multilingual IPA model training. 
Through training, the IPA model learns phonetic representations that are (quasi) language universal: fundamental speech sounds from  different languages that share the same IPA symbol are mapped to the same output neuron of the IPA model.
While any ASR architecture could be adopted in principle, we adopt the attention encoder-decoder E2E architecture, 
based on the finding in  \cite{zelasko2022discovering} -- it showed the superiority of the E2E over the hybrid architecture in multilingual phonetic sharing.  

The  IPA  model is trained with joint connectionist temporal classification (CTC) and attention objectives \cite{kim2017joint}. After training, 
we rely on the CTC posteriorgram built on top of the IPA model's encoder  to obtain   frame-aligned IPA label sequences of input speech utterances: every frame's label is determined by the output dimension having the  largest probability.
Given a set of $M$ distinct languages' untranscribed speech data (denoted as $\mathcal{G}$), their corresponding IPA pseudo labels are generated by the same IPA  model.
Notably, there is no assumption on the relation between $\mathcal{F}$ and $\mathcal{G}$:  In the case where there exists a speech sound in $\mathcal{G}$ which is not covered by the pseudo labeler's IPA inventories, the pseudo labeler will output an IPA symbol that best resembles the sound.


To address the concern of requiring supervised data that might be seemingly less attractive than  supervision-free pseudo labeling methods, we will  show 
in our experiments 
that IPA model training is \textit{not} data-hungry -- 100 hours per language suffice.

This approach is applicable to any languages, including those without knowledge of  grapheme-to-IPA conversion, because the  multilingual IPA model
can  estimate pseudo labels  that 
best describe  speech sounds of any languages.

\subsection{Phonetically-informed HuBERT model}

The HuBERT  model training comprises pretraining followed by finetuning. In the pretraining stage, HuBERT adopts BERT-like \cite{devlin2018bert} masked prediction. Offline K-means based speech frame clustering is carried out multiple rounds, in order to obtain refined K-means labels as the pretraining target.
Let $X$ and $Z$ denote a speech utterance and a  K-means label sequence respectively, of length $T$.  The masked prediction loss $L$ in a standard HuBERT is: 
\begin{equation}
    L = \alpha \cdot \sum_{t\in M}{\log p_f (z_t | \tilde{X}, t)} + (1-\alpha) \cdot \sum_{t  \notin  M}{\log p_f (z_t | \tilde{X}, t ) },
    \label{eqt:hubert_loss}
\end{equation}
where $M$ is a set of indices to be masked for $X$, $\tilde{X}$ is the masked variant of $X$ based on $M$. The parameter $\alpha$  is the  masked  prediction weight. 
The function $p_f$ denotes the prediction process which  outputs a distribution over the target $Z$. The function $p_f$ is parameterized using the HuBERT model's waveform encoder,  Transformer, codeword embedding layer and projection layer, as follows:
\begin{equation}
    p_f := \frac{\exp (\mathrm{sim} (\bm{A} \cdot \bm{o_t}, \bm{e_c} ) / \tau ) }{\sum_{c^{\prime}=1}^{C} \exp (\mathrm{sim} (\bm{A} \cdot \bm{o_t}, \bm{e_{c^{\prime}}})  / \tau )},
    \label{eqt:pf}
\end{equation}
where $\{\bm{o_t}|t\in 1 ... T\}$ is the Transformer output (Transformer takes as input from masked output of the waveform encoder), $\bm{e_c}$ is the embedding for codeword $c$, $\bm{A}$ is the projection matrix, $\mathrm{sim}(\cdot,\cdot)$ is the cosine similarity, $\tau$ is a scaler.

After pretraining, HuBERT finetuning is conducted using the  CTC loss  on a target language's ASR task. Finetuning is applied to the whole model except the waveform encoder. 

In this study, we improve the HuBERT pretraining stage, via  replacing the K-means clustering labels by IPA labels as the target of masked prediction. 
Guided by the language-universal and phonetically-rich IPA labels, 
the pretrained model benefits from multilingual phonetic sharing, which we believe is highly preferred in the subsequent low-resource ASR finetuning. 
Furthermore, having a well-trained multilingual IPA model, the IPA pseudo labeling process is much faster than the multi-round  clustering process   \cite{hsu2021hubert}.



\section{Experimental setup}
\subsection{Databases and evaluation metric}
\label{subsec:exp_databases}
Experiments are carried out using the open-source Multilingual Speech (MLS) corpus \cite{pratap2020mls}. The corpus covers Polish (PL), Portuguese (PT), Italian (IT), Spanish (SP), French (FR), Dutch (DU), German (GE) and English (EN), all derived from read audiobooks. 
The amount of training data per language
varies from 0.1k to 45k hours (see Table \ref{tab:database}). As this paper tackles  low-resource ASR, while 
the English training data size is 
too large to be considered ``low-resource'',
the English ASR task is excluded from this paper.

We consider two broad types of multilingual training sets for IPA model training and phonetically-informed HuBERT pretraining: (1) 
\textbf{MLS-7}: we merge  training sets of PL, PT, IT, SP, FR, DU and GE, summing up to \textbf{6.0k} hours. Besides, we   select  a random training subset of a certain size (at most) from each  language, and merge them as  limited-supervision versions of MLS-7. We construct \textbf{MLS-7-100h}, \textbf{MLS-7-250h} and \textbf{MLS-7-500h}, where the suffix denotes the   data size per language (if the full size of  a language's training set   is smaller than the suffix, we apply the full set). 
(2)  \textbf{MLS-8-en12kh}: 
    we select a random subset of 12k hours from the EN training set and merge with \textbf{MLS-7}, summing up to \textbf{18k} hours. We do not take the full EN training data as it would largely damage  the non-EN ASR tasks in our experiments.


For HuBERT finetuning,  one language's training set is used at a time. 
For each language we also prepare a random training subset of 100 hours or 500 hours (at most) for finetuning.
The finetuned ASR model is evaluated on a target language's test set, using word error rate (WER) as the metric.
The \textit{dev} data partition is used   to monitor the model training procedure.

\begin{table}[!t]
\renewcommand\arraystretch{0.7}
\centering
\caption{Speech data used in the experiments. Train, Dev and Test rows are presented in the number of hours.}
\resizebox{ 0.85 \linewidth}{!}{%
\begin{tabular}{c|cccccccc}      
\toprule
Language & PL & PT & IT & SP & FR & DU & GE & EN \\
\midrule
Train & 104 & 161  & 247  & 918  & 1.1k & 1.6k & 2.0k & 44.7k\\
Dev & 2.1 & 3.6 &5.2 & 10.0 & 10.1 &12.8 &14.3 & 15.8\\
Test & 2.1 &3.7 &5.3 &10.0 &10.1 &12.8 &14.3 &15.6 \\

\bottomrule
\end{tabular}%

}
\label{tab:database}
\end{table}

\subsection{Language-universal IPA model setup}
\label{subsec:setup_ipa_model}
The IPA model consists of a
Conformer \cite{guo2021recent_conformer} encoder and a Transformer \cite{karita2019comparative} decoder. It is implemented using ESPnet \cite{watanabe2018espnet}, trained with  IPA transcripts converted
using \cite{hasegawa2020grapheme}. Following \cite{Zelasko2020That_interspeech}, every IPA symbol, including the modifier (e.g. [\textlengthmark]), is modeled as a base unit by the IPA model.
The encoder and decoder have 18 layers and 2 layers respectively, with 4 attention heads,   768 attention dimensions and 2048 position-wise feed forward (PFF) dimensions. 
The Conformer encoder contains a 2-layer CNN with a kernel size of 31. Multiple  IPA models are trained, each  using  one of the multilingual datasets as mentioned in Section \ref{subsec:exp_databases}.  The decoder output dimension of the model, i.e. the IPA inventory size,  is 87 (training data excludes EN) or 95 (training data includes EN). 
The model is trained for 60 epochs  with joint CTC/attention objectives and the CTC weight is 0.1, using the  Adam optimizer \cite{kingma2014adam}, a peak learning rate of 0.001 and  2000 warm-up steps. The final model is obtained by taking average over models of the last 10 epochs.
\subsection{Phonetically-informed HuBERT setup}
\label{subsec:setup_hubert}
The HuBERT model is implemented by mainly following the setup from \cite{hsu2021hubert}. Two model sizes, i.e. `BASE' and `LARGE', are experimented. The models of both sizes contain a 7-layer 512-channel CNN encoder which outputs frames of 16kHz. The BASE-sized HuBERT contains a 12-layer Transformer with 8 attention heads, 768 attention dimensions and 3072 FFN dimensions, and a projection layer of 256 dimensions. For the LARGE-sized HuBERT, the numbers are 24, 16, 1024, 4096 and 768. Regarding masking, 8\% of the timesteps are randomly selected as starting positions, and spans of 10 steps are masked.
The HuBERT model is pretrained by  using one of the multilingual training sets in \{MLS-7, MLS-8-en12kh\} and corresponding IPA pseudo labels inferred from
one of the IPA models discussed in Section  \ref{subsec:setup_ipa_model}. 
The pretraining stops  after a fixed 400k iterations, using the Adam optimizer  \cite{kingma2014adam}, a peak learning rate of 5e-4 (BASE) or 1.5e-3 (LARGE), a warm-up step size of 32k. 
The parameter $\alpha$ is chosen from \{1, 0.5, 0.25, 0\} to determine the optimal value (see  Section \ref{subsec:results_effect_of_alpha}).

HuBERT finetuning is carried out on one of the 7 languages' training speech and orthographic transcripts (excluding English) at a time.
Our initial experiments revealed that the peak learning rate and the maximum iterations largely affect the  performance. Unless otherwise stated, we set the two values as 1.6e-4 and 150k. 
A freeze-step of 10k is set so that in the first 10k iterations, parameters of only the newly added softmax output layer are updated.
The final model selection after finetuning is based on the WERs on the dev set. A language model is \textit{not} used throughout this paper.




\subsection{Comparative methods}
\label{subsec:setup_compara_methods}
\subsubsection{Standard HuBERT}
We first compare our method with standard HuBERT models. We train   standard HuBERT models by strictly following the multi-round  K-means  label generation procedure stated in \cite{hsu2021hubert}. 
The 1st-round K-means   generates 100 clusters  on  39-dimensional MFCC features of the \textit{MLS-7-100h} set. 
The 1st-round K-means labels are used to pretrain a BASE-sized HuBERT model with the \textit{MLS-7} set (denoted as HB-Std-1). 
The 2nd-round  K-means   generates  500 clusters on features extracted from the 6-th Transformer layer output of HB-Std-1. The data for the 2nd-round clustering is a 10\% random subset of \textit{MLS-7-100h} due to the memory constraint.
The 2nd-round K-means labels are used to pretrain a BASE-sized HuBERT with the \textit{MLS-7} set, and is denoted as \textbf{HB-Std-2}. 
The 3rd-round K-means clustering is similar to the 2nd-round clustering except the speech features are extracted from the 9-th Transformer layer of HB-Std-2. 
The 3rd-round K-means labels are used to pretrain a LARGE-sized HuBERT with  \textit{MLS-7}, and is denoted as \textbf{HL-Std}. We take HB-Std-2 and HL-Std as two baseline pretrained models, and carry  out  finetuning and evaluation following the setup in Section \ref{subsec:setup_hubert}.


\subsubsection{IPA model retraining with BPE transcripts}
\label{subsubsec:setup_compara_method_bpe}
We then compare our method with an IPA model retraining method as proposed in our another paper \cite{Feng2023conformer}. 
We take the multilingual IPA model trained by \textit{MLS-7} as the seed model, keep its encoder meanwhile
replacing its decoder with a randomly initialized one,  and  retrain it with   a target language's training set using byte-pair encoding tokenized transcripts. 
The training setup follows Section \ref{subsec:setup_ipa_model}. We name this model as \textbf{nH-MLS-7}, where ``nH'' means non-HuBERT.

\section{Experimental results}

\subsection{Effect of masked prediction weight}
\label{subsec:results_effect_of_alpha}
This section shows the influence  of the masked prediction weight  $\alpha$ in phonetically-informed HuBERT pretraining. The comparison is made by taking the BASE-sized   architecture, using  \textit{MLS-7} as the pretraining data. The IPA pseudo labels are inferred by
the IPA model trained with \textit{MLS-7}.
The WER results are listed in Table \ref{tab:results_alpha}.

\begin{table}[!t]
\renewcommand\arraystretch{0.7}
\centering
\caption{WER$\%$ results w.r.t different values of $\alpha$. The HuBERT model is BASE-sized, pretrained with \textit{MLS-7}. The IPA model for pseudo labeling was trained by \textit{MLS-7}.}
\resizebox{ 0.8 \linewidth}{!}{%
\begin{tabular}{l|ccccccc|c}      
\toprule
$\alpha$ & PL & PT & IT & SP & FR & DU & GE & Avg. \\
\midrule
1.0 & 19.42  & 22.98 & 14.59 & 9.61 & 10.02 & 17.20 & 9.86 & 14.81 \\
0.5 & 12.35 & 21.61 & 13.77 & 7.37 & 9.07 & 14.24 & 9.40 & \textbf{12.54} \\
0.25 &  14.17 & 22.86 & 14.28 & 8.23 & 9.61 & 14.42 & 9.84 & 13.34 \\
0 & 17.46 &25.30 &15.83 &9.97 &11.29 &16.33&11.28 &15.35 \\

\bottomrule
\end{tabular}%
}
\label{tab:results_alpha}
\end{table}

The proposed approach performs the best when $\alpha$ is set to 0.5. This indicates in the proposed phonetically-informed pretraining, the HuBERT model benefits from joint masked and unmasked prediction. 
When $\alpha=1.0$, i.e. masked prediction only, our approach degrades by an absolute 2.3\% compared to $\alpha=0.5$, which is opposed to the finding in   K-means based pretraining  \cite{hsu2021hubert}. 
We  conclude that while masked prediction is crucial in pretraining   irrespective of the use of  IPA  or K-means pseudo labels, when phonetically-driven labels are leveraged, unmasked prediction also plays an important role. Setting $\alpha=0$ results in the worst performance,
 similar to \cite{hsu2021hubert}. 
 For the rest of this section, $\alpha$ is fixed as 0.5.




\subsection{Training an IPA model is not data-hungry}
This section probes  the effect of the language-universal IPA model  (i.e. the pseudo labeler) in phonetically-informed HuBERT pretraining.
Four IPA models are compared, each trained with one of the MLS-7, MLS-7-100h, MLS-7-250h and MLS-7-500h sets. 
Each  IPA model creates pseudo labels for MLS-7 which are then used to pretrain a  HuBERT model. 
WER\% results of finetuned HuBERT models  are shown in Table \ref{tab:results_ipa_model_train_amount}. 
When the HuBERT   size   is BASE, reducing the training data amount for the IPA model from MLS-7 (6.0k hours) to MLS-7-100h (0.7k hours) leads to negligible   degradation.
When the HuBERT   size is LARGE, the absolute WER degradation is 0.58\%.
Table \ref{tab:results_ipa_model_train_amount} also shows that the proposed approach, irrespective of using full or part of the data in training an IPA model, consistently outperforms the standard HuBERT models (HB-Std-2 and HL-Std), and this is true for both BASE and LARGE sizes.  

To gain a deeper understanding on the quality of IPA pseudo labels, we conduct an ablation study, i.e. phone recognition by the above-mentioned IPA models. The average phone token error rate (PTER) results over the 7 languages' test sets are reported in Table \ref{tab:results_ablation_phone_recognition}. Taking Tables \ref{tab:results_ablation_phone_recognition}  and \ref{tab:results_ipa_model_train_amount}, reducing IPA model's training data has a much greater negative impact on phone recognition  (1.8\% absolutely) than that on phonetically-informed speech pretraining (0.08\% BASE/0.58\% LARGE). 
This may be  explained that
some inaccurate pseudo labels are consistent over frames, which may still provide useful guidance for   pretraining. 
Conclusively, training an IPA model for the proposed approach is not data-hungry.

\begin{table}[!t]
\renewcommand\arraystretch{0.5}
\centering
\caption{WER$\%$ of HuBERT models with different IPA pseudo labels and baseline   with K-means labels. 
Models were pretrained with \textit{MLS-7}. ``$*$'' denotes the language of which the IPA model training amount is less than that claimed (suffix). }
\resizebox{ 0.99 \linewidth}{!}{%
\begin{tabular}{l|lllllll|c}      
\toprule
Train set for IPA model  & PL & PT & IT & SP & FR & DU & GE & Avg. \\
\midrule
\multicolumn{9}{l}{HuBERT Transformer size: BASE} \\ 
\midrule
 MLS-7& 12.35 & 21.61 & 13.77 & 7.37 & 9.07 & 14.24 & 9.40 & 12.54 \\
 MLS-7-500h& 12.47$^*$ & 20.90$^*$ & 13.22$^*$ & 8.47 & 9.11 & 14.06 & 9.42 & 12.52 \\
 MLS-7-250h&  13.52$^*$ & 20.30$^*$ & 13.16$^*$ & 8.21 & 9.05 & 14.06 & 9.44 & 12.53 \\
 MLS-7-100h & 11.64 & 22.25 & 13.53 & 7.86 & 9.11 & 14.18 & 9.75 & 12.62 \\
 HB-Std-2 (baseline) &  17.91 & 21.73 & 16.22 & 7.67 & 10.46 & 15.22 & 11.74 & 14.42  \\
 \midrule
\multicolumn{9}{l}{HuBERT Transformer size: LARGE} \\ 
\midrule
MLS-7 & 5.35 & 13.22 & 10.00 & 5.49 & 6.58 & 11.47 & 7.77 & 8.55\\
 MLS-7-100h & 6.38 & 14.38 & 10.65 & 5.51 & 7.11 & 11.86 & 8.02 & 9.13 \\
 HL-Std (baseline) & 6.81 & 16.26 & 10.64 & 6.05 & 7.33 & 11.57 & 8.78 & 9.63 \\

\bottomrule
\end{tabular}%
}
\label{tab:results_ipa_model_train_amount}
\end{table}

\begin{table}[!t]
\renewcommand\arraystretch{0.5}
\centering
\caption{Phone token error rate $\%$ by different IPA models. The  result is averaged over the 7 languages' test sets.}
\resizebox{ 0.9 \linewidth}{!}{%
\begin{tabular}{l|cccc}      
\toprule
Train set for IPA model & MLS-7 & MLS-7-500h & MLS-7-250h & MLS-7-100h \\
\midrule
Avg. PTER\% & 4.16 & 4.60 & 4.83 & 5.96\\



\bottomrule
\end{tabular}%
}
\label{tab:results_ablation_phone_recognition}
\end{table}

\subsection{Comparison to other methods}
The comparison of the proposed approach, recent state of the arts (SotA) and the comparative method \textit{nH-MLS-7} \cite{Feng2023conformer} (see Section \ref{subsec:setup_compara_methods})  are summarized in Table \ref{tab:results_compare_to_other_methods}. The other comparative method, i.e. the standard HuBERT, is included in Table \ref{tab:results_ipa_model_train_amount}.
The model's number of parameters  is enclosed in brackets.
In all the XLSR-53  \cite{conneau2020unsupervised}, B0 (15-lang init.) \cite{li2021scaling} and JUST \cite{bai2022joint} models, the 8 MLS language's training data is fully used.
In the proposed approach, we find it detrimental  by including the full   EN training set  during pretraining: taking   LARGE,  the averaged WER is 11.13\% (not listed in the table) by merging EN and MLS-7, and is  8.55\% by using MLS-7 only. 
We find it beneficial by including a 12k-hour subset of the  EN  set (i.e. \textit{MLS-8-en12kh}) on the BASE-sized model pretraining. Although   MLS-8-en12kh does not improve the LARGE-sized model compared to MLS-7, we report results of both sizes in Table \ref{tab:results_compare_to_other_methods}. 

From Table \ref{tab:results_compare_to_other_methods}, our approach ``MLS-7'' with  the LARGE  size performs better than XLSR-53 and B0 (15-lang init.) on average, and utilize  much less pretraining data in terms of   hours and language diversity. Compared to XLSR-53, our approach consumes additional supervision information in pretraining. Looking at ``MLS-7-100h'' in Table \ref{tab:results_ipa_model_train_amount}, with only 100-hour per language's supervision, the proposed approach  starts to outperform XLSR-53. Moreover, our systems do not adopt a LM while XLSR-53 adopts one. As for B0 (15 lang init.), it consumes supervised training data of 359k hours covering 15 languages to pretrain a seed encoder.
The JUST model performs better than our approach. JUST uses more training   resources and a slightly larger model architecture. Lastly, our approach (LARGE) performs better than \textit{nH-MLS-7}. 


\begin{table}[!t]
\renewcommand\arraystretch{0.5}
\centering
\caption{WER$\%$ of SotA works, comparative methods and the proposed approach. The column ``Hrs'' denotes the amount of pretraining data in hours. The number enclosed in brackets denotes the parameter number. }
\resizebox{ 0.99 \linewidth}{!}{%
\begin{tabular}{l|l|lllllll|c}      
\toprule
 &Hrs & PL & PT & IT & SP & FR & DU & GE & Avg. \\
 \midrule
 XLSR-53 (300M) \cite{conneau2020unsupervised} & 56k & 17.2&14.7&10.4&6.3&7.6&10.8&7.0&10.6 \\
 B0 (15-lang init.; 370M) \cite{li2021scaling} & 359k & 10.9 & 15.5 & 10.1 & 4.7 & 6.1 & 11.1 & 5.0 & 9.1 \\
 JUST (600M) \cite{bai2022joint} & 50k & 6.6& 8.0& 8.2& 3.7& 5.2& 9.5& 4.1& 6.5 \\ 
 \midrule
 nH-MLS-7 (216M) \cite{Feng2023conformer} & 6k & 14.14 &14.84 &10.89 &5.36 &4.97 &11.37 &6.90 &9.78\\
\midrule
\multicolumn{10}{l}{HuBERT Transformer size: LARGE (317M)} \\
\midrule
MLS-7 & 6k & 5.35 & 13.22 & 10.00 & 5.49 & 6.58 & 11.47 & 7.77 & 8.55\\
MLS-8-en12kh &18k & 10.50& 17.25& 11.70& 6.44& 6.22& 11.71& 7.63& 10.21\\
 \midrule
 \multicolumn{10}{l}{HuBERT Transformer size: BASE (95M)} \\
 \midrule
 MLS-7& 6k & 12.35 & 21.61 & 13.77 & 7.37 & 9.07 & 14.24 & 9.40 & 12.54 \\
 MLS-8-en12kh &18k &10.39& 22.38& 12.42& 8.98& 8.63& 13.79& 9.23& 12.26  \\

\bottomrule
\end{tabular}%
}
\label{tab:results_compare_to_other_methods}
\end{table}

\subsection{Robustness towards limited finetuning data}
WER\% results of the proposed approach and comparative methods using limited finetuning (FT) data are summarized in Table \ref{tab:results_ft_amount}. 
Notably, all the proposed systems in Table \ref{tab:results_ft_amount} use  MLS-7 for speech pretraining, and the first column  denotes the training set for the IPA model.
We have the following observations: 

(1) Compared to standard HuBERT models, our phonetically-informed HuBERT is able to perform better meanwhile consuming   less supervised data. For instance, on German, the WER of   ``HL-Std'' (in Table \ref{tab:results_ipa_model_train_amount}) is 8.78\% by adopting the full  FT data (2k hours), whereas the WER of the LARGE-sized ``MLS-7-100h'' is 8.55\% by adopting only  500 hours of FT data plus  100 hours of supervised data for IPA model training. The 100-hour set is a subset of the 500-hour FT set, so our approach saves   1.5k hours (i.e. 75\%) of German supervised data. such observations are   also
seen on French, Spanish (LARGE) and German, Dutch,  French (BASE).

(2) Compared to XLSR-53 and nH-MLS-7, the proposed LARGE-sized MLS-7 system achieves better performances in the  100-hour FT data setting. Moreover, our system has a smaller WER degradation (absolute 0.56\%) when the FT data reduced from full to 100 hours, comparing to XLSR-53 (1.2\%) and nH-MLS-7 (2.6\%).


\begin{table}[!t]
\renewcommand\arraystretch{0.5}
\centering
\caption{WER$\%$ of the proposed approach and comparative methods w.r.t. limited finetuning (FT) data.  ``*'' denotes a language of which the full data amount is less than 500 hours. All the proposed models used MLS-7 for pretraining. The first column indicates the training set for training the IPA model. }
\resizebox{ 0.99 \linewidth}{!}{%
\begin{tabular}{l|l|lllllll|c}      
\toprule
 & FT data in hrs  & PL & PT & IT & SP & FR & DU & GE & Avg. \\
\midrule
XLSR-53 \cite{conneau2020unsupervised} & 100 & 18.9& 15.7& 12.0& 7.9& 9.8& 10.9& 7.4& 11.8 \\ 
nH-MLS-7 & $100$ &16.41& 15.54& 12.00& 7.97& 8.62& 15.62& 10.45& 12.37\\
\midrule
\multicolumn{10}{l}{HuBERT Transformer size: LARGE} \\
\midrule
MLS-7-100h & $\min\{500,\textrm{full}\}$& 6.38$^*$& 14.38$^*$& 10.65$^*$& 5.52& 7.29& 12.72& 8.55& 9.36 \\
MLS-7 & $100$ &5.35& 13.12& 10.60& 5.92& 7.57& 12.64& 8.54& 9.11 \\
\midrule
\multicolumn{10}{l}{HuBERT Transformer size: BASE} \\
\midrule
MLS-7-100h & $\min\{500,\textrm{full}\}$ & 11.64$^*$& 22.25$^*$& 13.53$^*$& 8.06& 9.39& 15.22& 10.57& 12.95\\
MLS-7 & $100$ & 11.10& 22.79& 15.31& 9.98& 12.04& 17.55& 12.30& 14.44 \\

\bottomrule
\end{tabular}%
}
\label{tab:results_ft_amount}
\end{table}
\section{Conclusions}
This paper proposed a speech pretraining approach to improving the HuBERT  model for low-resource ASR. At its core, language-universal phonetic pseudo  labels inferred from a multilingual IPA model guide the speech pretraining process.
Experiments on the MLS corpus have shown that our approach consistently outperforms the standard HuBERT model over all the target languages, and on both model sizes. On German, Spanish, Dutch and French, our approach  performed better than (or similar to) the standard HuBERT,  and was able to save supervised data by 1.5k hours at most.
Our approach, consuming much less pretraining data than   XLSR-53 \cite{conneau2020unsupervised} and B0 (15-lang init.) \cite{li2021scaling}, performed better than both of them in terms of the averaged WER. Our approach also  outperformed a model retrained from  a multilingual IPA model.
Compared to XLSR-53 and a model retrained from a multilingual IPA model, our approach was more robust to limited finetuning data scenarios.





\bibliographystyle{IEEEtran}
\bibliography{mybib}

\end{document}